\documentclass[twocolumn,superscriptaddress,amsmath,amssymb,aps,prl]{revtex4-2}

\usepackage{graphicx}
\usepackage{dcolumn}
\usepackage{bm}
\usepackage{amsthm}
\newtheorem*{theorem*}{Theorem}
\usepackage{braket}

\usepackage[colorlinks,urlcolor=blue,citecolor=blue,linkcolor=blue]{hyperref}

\begin{document}
	\title{Synthetic Mutual Gauge Field in Microwave-Shielded Polar Molecular Gases}

	\author{Bei Xu}
	\affiliation{Institute for Advanced Study, Tsinghua University, Beijing, 100084, China}
	\author{Fan Yang}
	\affiliation{School of Physics, Renmin University of China, Beijing, 100872, China}
	\affiliation{Key Laboratory of Quantum State Construction and Manipulation (Ministry of Education), Renmin University of China, Beijing, 100872, China}
	\author{Ran Qi}
	\affiliation{School of Physics, Renmin University of China, Beijing, 100872, China}
	\affiliation{Key Laboratory of Quantum State Construction and Manipulation (Ministry of Education), Renmin University of China, Beijing, 100872, China}
	\author{Hui Zhai}
	\email{hzhai@tsinghua.edu.cn}
	\affiliation{Institute for Advanced Study, Tsinghua University, Beijing, 100084, China}
	\affiliation{Hefei National Laboratory, Hefei 230088, China}
	\author{Peng Zhang}
	\email{pengzhang@ruc.edu.cn}
	\affiliation{School of Physics, Renmin University of China, Beijing, 100872, China}
	\affiliation{Key Laboratory of Quantum State Construction and Manipulation (Ministry of Education), Renmin University of China, Beijing, 100872, China}
	\date{\today}
	
	\begin{abstract}
The recent breakthrough of realizing the Bose-Einstein condensate of polar molecules and degenerate Fermi molecules in three dimensions relies crucially on the microwave shielding technique, which strongly suppresses the collision loss between molecules. In this letter, we show that the cooperation of microwave shielding and dipolar interaction naturally leads to the emergence of a synthetic gauge field. Unlike that studied in cold atoms before, this gauge field couples to the relative motion of every two molecules instead of single-particle motion, therefore being a mutual gauge field. In this case, every molecule carrying a synthetic charge sees the other molecule as carrying the source of the magnetic field, and the spatial distribution of the magnetic field is reminiscent of a solenoid attached to the molecule. In other words, in addition to microwave-shielded interaction, another part of the interaction between two molecules behaves as a charge interacting with a solenoid, which was missed in the previous discussion. We argue that the physical manifestation of this gauge field is breaking time-reversal symmetry in the collective spatial motion of molecules. Finally, we discuss the challenges in quantitatively studying such a quantum many-body system. 

	\end{abstract}

	\maketitle
	
Since achieving Bose-Einstein condensation of atoms, it has been a long-lasting effort for over two decades to realize quantum degenerate gases of polar molecules, which offer the promise of richer quantum many-body phenomena thanks to their long-range dipolar interactions~\cite{Carr2009,Bohn2017,Cornish2024,Langen2024}. However, the main obstacle is the sticky collision loss between molecules because of their complicated short-range interaction potentials~\cite{Bause2023,Langen2024}. When these molecules are confined in two dimensions, a DC electric field is sufficient to reduce the collision loss because the dipolar interaction is repulsive with these molecules' dipole moments polarized perpendicular to the plane~\cite{Valtolina2020}. Nevertheless, in three dimensions, a DC electric field cannot make interaction repulsive in all directions and cannot prevent collision loss. Later, it was proposed that a circularly polarized microwave coupling different molecular rotational states can engineer a three-dimensional repulsive short-range potential to suppress collision loss, a technique called microwave shielding~\cite{Karman2018, Lassabliere2018}. Since 2021, several experimental groups have realized the suppression of collision loss by microwave shielding in optical tweezers and traps~\cite{Anderegg2021, Schindewolf2022,Lin2023, Bigagli2023,Chen2023,Chen2024,molecularBEC2024}. With this new technique, fermionic molecules have been cooled to below the Fermi temperature~\cite{Schindewolf2022}, and Bose-Einstein condensation of bosonic molecules~\cite{molecularBEC2024} has been recently realized.  

\begin{figure}[t!]
    \centering
    \includegraphics[width=0.48\textwidth]{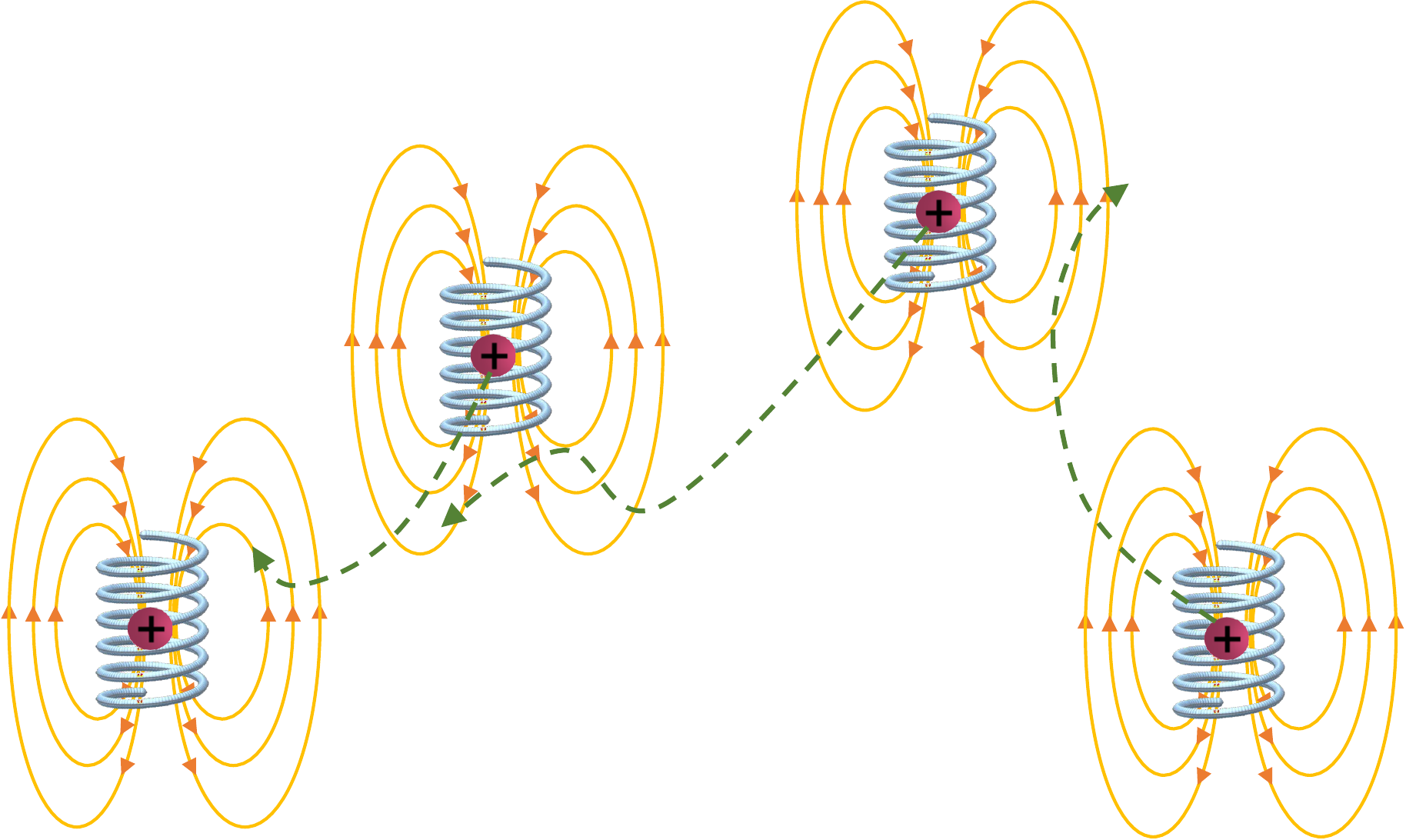}
    \caption{Schematic of synthetic mutual gauge field as solenoid attachment. Each molecule carrying a synthetic charge (denoted as red balls) sees a synthetic magnetic field produced by other molecules (denoted by yellow lines with arrows), and the synthetic magnetic field looks like a solenoid is attached to these molecules. The synthetic magnetic field bends the spatial trajectories of these molecules (dashed lines).}
    \label{mutual-gauge}
\end{figure}

In this letter, we point out that microwave shielding is not only a tool to overcome the obstacle, but it can also generate an intriguing correlated quantum many-body system that has never been encountered or discussed before. In short, microwave shielding, cooperating with dipolar interactions between molecules, generates a synthetic gauge field in this system, which possesses the following two unconventional features:

First, the synthetic gauge field has been extensively investigated in cold atom physics before, where such a gauge field is coupled to the motion of each single particle~\cite{Lin2009,Dalibard2011,Lin2011,Wang2012,Cheuk2012,Goldman2014,Zhai2015,Huang2016,Wu2016,Cooper2019,Wang2021}. In our case, we show that, in a two-molecule system, the gauge field emerges and couples to the relative coordinate of two molecules. That is to say, one molecule sees another molecule as a moving source of the synthetic magnetic field. In a many-body system, every molecule carries a synthetic charge and sees any other molecule as carrying the source of the magnetic field. This is reminiscent of the composite fermion description of the fraction quantum Hall effect, where such a gauge field is called the mutual gauge field~\cite{Nagaosa}. 

Secondly, the mutual gauge field manifests as flux attachment in the fractional quantum Hall case. Namely, a magnetic field is attached to each electron. However, each electron does not see the magnetic field attached to itself but sees the magnetic flux attached to other electrons. In a two-molecule system, the synthetic mutual gauge field behaves like \textit{solenoid attachment} instead of flux attachment. That is to say, one molecule behaving as a charged particle sees the other molecule as if a solenoid is attached to it. More concretely, in the relative coordinate, the synthetic magnetic field lines go from the north pole to the south pole, where the microwave polarization defines the poles. We emphasize that this is not the scenario in which two molecules interact by magnetic dipolar interaction as two solenoids.

In Fig. \ref{mutual-gauge}, we schematically illustrate this physical picture of the mutual synthetic gauge field in microwave-shielded polar molecules. Below, we will first show how such a gauge field emerges and then discuss its physical consequences. 

\begin{figure}[t!]
    \centering
    \includegraphics[width=0.48\textwidth]{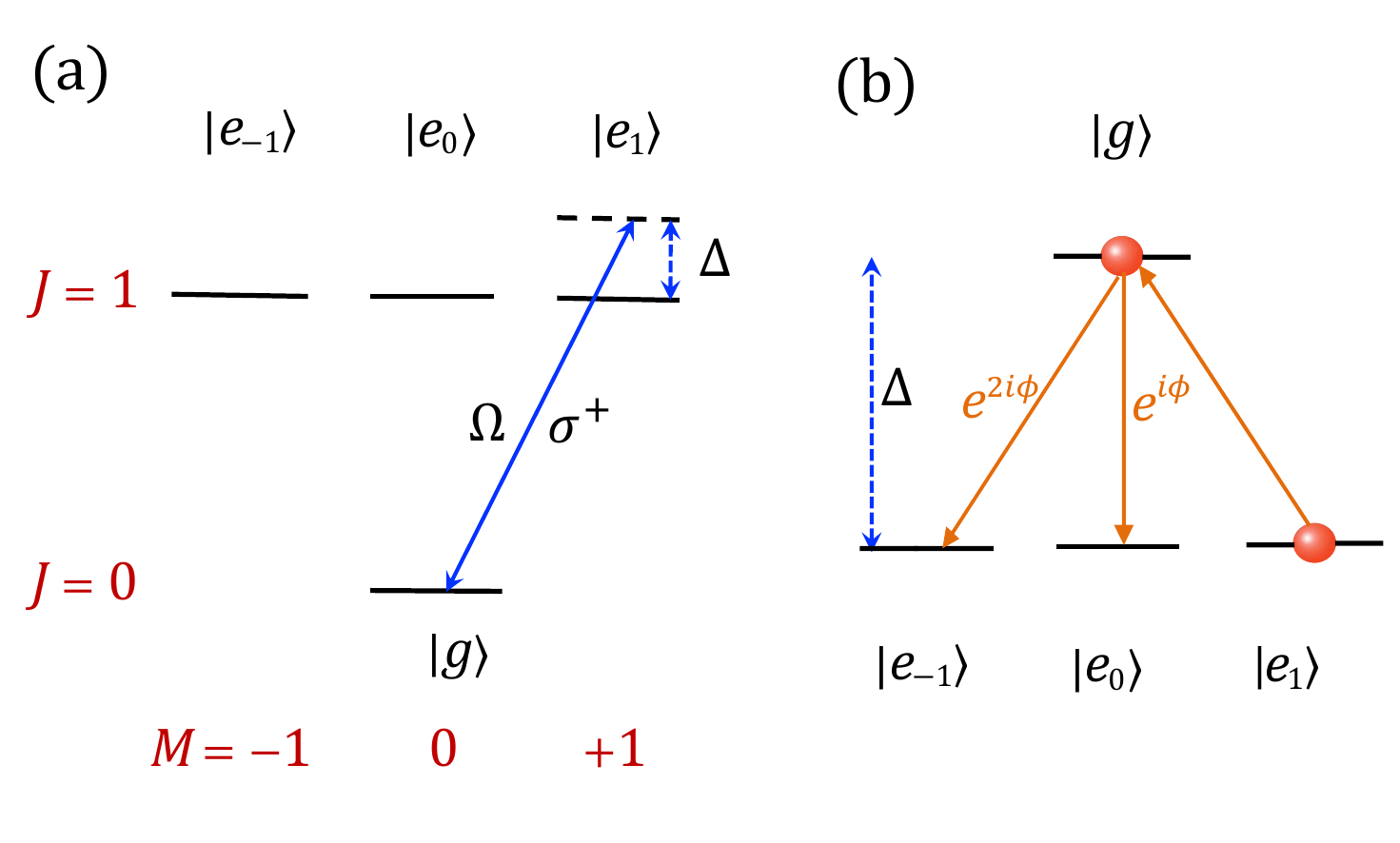}
    \caption{{\bf (a):} Single-particle Hamiltonian in the lab frame. For each molecule, among the four internal rotational states, $\ket{g}$ and $\ket{e_1}$ are coupled by circular polarized microwave $\sigma^+$ with strength $\Omega$. {\bf (b):} Dipolar interactions in the rotating frame. The arrows denote two typical interaction processes and the phase factors attached to these processes.}
    \label{system}
\end{figure}

\textit{Physical System.} For each molecule, four internal rotational states with angular momentum $J=0$ and $J=1$ are mostly relevant to this discussion, and we denote them by $\ket{g}\equiv\ket{0,0}$, $\ket{e_{-1}}\equiv\ket{1,-1}$, $\ket{e_0}\equiv\ket{1,0}$ 
 and $\ket{e_1}\equiv\ket{1,1}$, where $|J,M\rangle$ is an eigen-state with  quantum number $J$ for internal angular momentum  and $M$ for its $z$-component. As shown in Fig. \ref{system}(a), we consider a $\sigma_+$-polarized microwave beam that couples $\ket{g}$ to $\ket{e_1}$. The single-particle Hamiltonian reads
\begin{eqnarray}
\hat H_0=2 \Lambda \sum_{s=0,\pm 1} \ket{e_s}\bra{e_s}+{\Omega}\cos(\omega t)\ket{e_1}\bra{g}+h.c.,\label{hsj}
\end{eqnarray}
 where $\Lambda$ is the molecular rotational constant, and $\Omega$ and $\omega$ are the coupling intensity and 
 angular
 frequency of the $\sigma^+$ beam, respectively. The dipolar interaction $\hat{H}_\text{int}$ between two polar molecules reads 
 \begin{eqnarray}
\hat H_{\rm int}({\bm r})=\frac{1}{4\pi \epsilon_0 r^3}
\bigg[{\hat {\bm{d}}}_1 \cdot {\hat {\bm{d}}}_2-\frac 3{r^2}({\hat {\bm{d}}}_1 \cdot \bm{r})({\hat {\bm{d}}}_2 \cdot \bm{r})\bigg],\label{vsdd}
\end{eqnarray}
where ${\bm r}\equiv {\bm r}_1-{\bm r}_2$ is the relative position of the two molecules, ${\hat {\bm{d}}}_j$ ($j=1,2$) is the electric dipole operator 
of the molecule $j$, and $\epsilon_0$ is the dielectric constant. The strength of the dipolar interaction can also be characterized by a bare dipolar length defined as $a_\text{d}=\mu d^2/(6\pi\hbar^2\epsilon_0 )$, with $\mu$ being the reduced mass and $d$ being the dipole moment. For example, for NaCs molecule, we have $a_\text{d}\approx 3.3\times 10^5a_0$, where $a_0$ is the Bohr radius.

We apply the rotating-wave approximation to eliminate the largest energy scale $\Lambda$, and in the rotating frame, $\hat{H}_0$ becomes 
\begin{eqnarray}
\hat H_0=\Delta \ket{g}\bra{g}+\frac{\Omega}{2}\ket{e_1}\bra{g}+h.c.,\label{hsj}
\end{eqnarray}
and $\Delta=\omega-2\Lambda$. We focus on the blue detuning case with $\Delta>0$ for discussing the microwave shielding. We project the dipole interaction into the subspace spanned by these four internal states of each molecule. With the rotating-wave approximation, we only keep the near-resonance interaction processes. Then, in the rotating frame, the dipolar interaction becomes~\cite{Deng2023}
    \begin{equation}
\hat H_\text{int}({\bm r})
=-\frac{2\sqrt{2\pi}d^2}{\sqrt{15}\epsilon_0 r^3}\sum_{m=-2}^{2}
\bigg[ \hat \Sigma_{2,m}{Y}^{m}_{2}(\theta,\phi)^\ast\bigg].\label{vdd}
\end{equation}
Here $\theta$ and $\phi$ are the polar and the azimuthal angles of ${\bm r}$, respectively, and
 ${Y}^{m}_{2}(\theta,\phi)$ are the spherical harmonics. The operators $\hat \Sigma_{2,m}$ ($m=\pm 2, \pm 1, 0$) are given by
  \begin{eqnarray}
{\hat \Sigma}_{2,0}&=&\frac{1}{4\pi\sqrt{6}}\big(2\ket{e_0;g}\bra{g;e_0}
-\ket{e_1;g}\bra{g;e_1}\big.\nonumber\\
&&\ \ \ \ \ \ \ \ \ \big.-\ket{e_{-1};g}\bra{g;e_{-1}}+\text{h.c.}\big), \label{s20}\\
{\hat \Sigma}_{2,1}&=&\frac{1}{4\pi\sqrt{2}}\big(|e_1;g\rangle\langle g; e_0|
-|e_0;g\rangle\langle g; e_{-1}|\big.\nonumber\\
&&\ \ \ \ \ \ \ \ \ \big.+|g;e_1\rangle\langle e_0;g|-|g;e_{0}\rangle\langle e_{-1}; g|\big), \label{s21}\\
{\hat \Sigma}_{2,2}&=&-\frac{1}{4\pi}\big(
|e_1;g\rangle\langle g; e_{-1}|+|g; e_1\rangle\langle e_{-1};g| \big), \label{s22}
\end{eqnarray}
and ${\hat \Sigma}_{2,-1}=-\Sigma_{2,1}^{\dagger}$ and ${\hat \Sigma}_{2,-2}={\hat \Sigma}_{2,2}^\dagger$,
where $|\alpha_1;\alpha_2\rangle\equiv|\alpha_1\rangle_1|\alpha_2\rangle_2$ $(\alpha_{1,2}=g,e_{0,\pm 1})$ is the direct product state of the two molecules. 
 Here $\hat{\Sigma}_{2,0}$ process conserves total $M$. $\hat{\Sigma}_{2,\pm 1}$ and  $\hat{\Sigma}_{2,\pm 2}$ processes change total $M$ by one and two, respectively, and therefore, these two processes are associated with a phase factor $e^{\mp i\phi}$ in $Y_{2}^{\pm 1}(\theta,\phi)^\ast$ and $e^{\mp 2i\phi}$ in $Y_{2}^{\pm 2}(\theta,\phi)^\ast$, respectively. These processes are shown in Fig. \ref{system}(b). These phase factors include all complex numbers in the Hamiltonian after the rotating-wave approximation.

\begin{figure}[t!]
    \centering
    \includegraphics[width=0.42\textwidth]{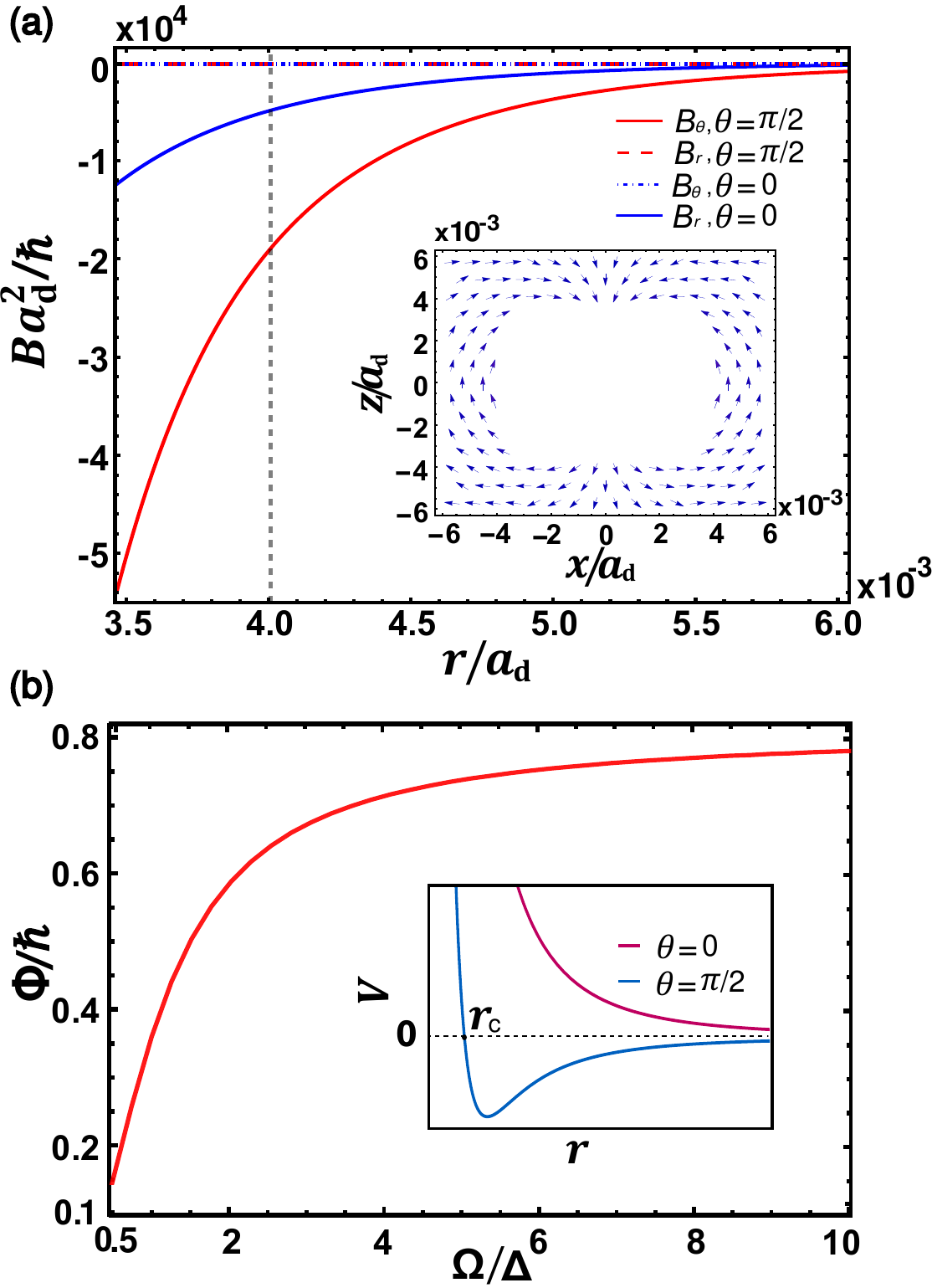}
    \caption{{\bf (a):} The synthetic magnetic field.
    The two solid lines are $B_r(r, \theta= 0)$ and $B_\theta(r, \theta=\pi/2)$, respectively, and the other two dashed lines with vanishing values are  $B_r(r, \theta=\pi/2)$ and $B_\theta(r, \theta= 0)$, respectively. The gray vertical dashed line indicates the location of the shielding core $r_\text{c}$, defined as the zero-crossing of $V(r,\theta=\pi/2)$. $a_d$ is taken as the length unit, of the order of $10^5a_0$. The inset shows a magnetic field distribution along the $xz$ plane outside of $r_\text{c}$. 
    {\bf (b):} Total magnetic flux 
 across the equator in the regime outside of the shielding core $r_\text{c}$. Here, the total flux is plotted as a function of $\Omega/\Delta$. The inset illustrates the microwave-shielded potential and the definition of $r_\text{c}$. Here, we take 
NaCs molecule as an example, and we set $\Delta=\hbar(2\pi)10{\rm MHz}$. For (a), we take $\Omega=\Delta$.  }
    \label{gauge}
\end{figure}

\textit{Synthetic Gauge Field.} We consider a two-body problem with $j=1,2$ denoting two molecules. The total Hamiltonian in the center-of-mass framework reads
\begin{align}
&\hat{\mathcal{H}}=-\frac{\hbar^2\nabla^2_{\bm r}}{2\mu}+\hat{W}({\bm r}), \\
&\hat{W}({\bm r})=\sum\limits_{j=1,2}\hat{H}^{(j)}_0+\hat{H}_\text{int}({\bm r}),
\end{align}
in the rotating frame,
where
$\hat{H}^{(j)}_0$ is $\hat H_0$ of the molecule $j$, and
 $\hat{W}({\bm r})$ describes the internal couplings between different rotational states due to both microwave shielding and dipolar interactions. 

The phases attached to the scattering processes shown in Fig. \ref{system} suggest that a unitary transformation $\hat{U}(\phi)$ can make $\hat{\widetilde{W}}(r,\theta)=\hat{U}^\dag(\phi)\hat{W}({\bm r})\hat{U}(\phi)$ as a $\phi$-independent and real one, where $\hat{U}(\phi)$ is defined as
\begin{eqnarray}
{\hat U}(\phi)=\exp[i \phi
(2\hat N_{e_{-1}}+\hat N_{e_0}
)],\label{up}
\end{eqnarray}
and $\hat N_{e_{i}}$ ($i=0,\pm 1$) counts total number of molecules in $\ket{e_{i}}$ state. Thus, $\hat{\widetilde{W}}(r,\theta)$ shares the same set of eigenvalues as $\hat{W}(r,\theta)$. By diagonalizing $\hat{\widetilde{W}}(r,\theta)$, we can obtain the highest eigenvalue, denoted by $V(r,\theta)$,  and its corresponding real eigenstate $\ket{\zeta(r,\theta}$. Here $V(r,\theta)$ 
is the microwave-shielded potential~\cite{Karman2018, Lassabliere2018,Deng2023}. It is an anti-dipolar potential, which is purely repulsive along $\theta=0$ and has a zero-crossing for $\theta=\pi/2$, as shown in the inset of Fig. \ref{gauge}. We define the zero-crossing $r_\text{c}$ as the shielding core because the repulsion of $V(r,\theta)$ increases rapidly for all directions when $r<r_\text{c}$. It forbids two molecules from being closer, which is also why microwave shielding can reduce collision loss. Typically, $r_\text{c}$ is of the order of $10^{-3}$ of $a_d$  and decreases smoothly as $\Omega/\Delta$ increases.

 The separation of different eigenvalues of $\hat{\widetilde{W}}(r,\theta)$ is large enough to ensure the adiabatic approximation, as justified by recent experiments~\cite{Anderegg2021, Schindewolf2022,Lin2023, Bigagli2023,Chen2023,Chen2024,molecularBEC2024}, the internal state follows the eigenstate $\hat{U}(\phi)\ket{\zeta(r,\theta}$ of $\hat{W}(r,\theta)$, and thus the total wave function of the two-molecule system in the  center-of-mass framework is given by
\begin{equation}
|\Psi({\bm r})\rangle=\psi({\bm r})\hat U(\phi)|\zeta(r,\theta)\rangle.\label{Psi}
\end{equation}
Here $\psi({\bm r})$ experiences an effective 
Hamiltonian~\cite{AldenMead1980,geo1989,sun1990}
\begin{equation}
\hat{\mathcal{H}}=\frac{1}{2\mu}\left[-i\hbar\nabla_{{\bm r}}-{\bm A}({\bm r})\right]^2+V(r,\theta),
\end{equation}
where ${\bm A}({\bm r})=i\hbar\bra{\zeta(r,\theta)}\hat{U}^\dag\nabla_{\bm r}[\hat{U}\ket{\zeta(r,\theta)}]$ is the induced gauge field.
Since  $\ket{\zeta(r,\theta}$ is real and $\hat{U}$ only depends on $\phi$, ${\bm A}$ only has ${\bm e}_\phi$ component as ${\bm A}({\bm r})=A_\phi(r,\theta){\bm e}_\phi$, and 
\begin{equation}
A_\phi(r,\theta)=\frac{-\hbar}{r\sin\theta}\bra {\zeta(r,\theta)}\big(2\hat N_{e_{-1}}+\hat N_{e_0}\big)\ket{\zeta(r,\theta)}. 
\end{equation}
We can further derive the synthetic magnetic field as ${\bm B}({\bm r})=\nabla\times {\bm A}({\bm r})=B_r(r,\theta){\bm e}_r+B_\theta(r,\theta){\bm e}_\theta$. This synthetic magnetic field has no $\phi$ dependence, respecting the cylindrical symmetry of $\sigma^+$ microwave. 

We show a numerical illustration of  ${\bm B}({\bm r})$ in Fig. \ref{gauge}. First, in the inset of Fig. \ref{gauge}(a), we plot the synthetic magnetic field distribution in the $xz$ plane outside of the shielding core, which clearly shows that the magnetic field distribution follows the distribution produced by a solenoid outside of the shielding core $r_\text{c}$. Following the cylindrical symmetry, rotating this plot along $\hat{z}$ generates the magnetic field distribution in the entire three-dimensional space. In Fig. \ref{gauge}(a), we also show that at $\theta=0$, ${\bm B}$ is purely along the radial direction. At the equator with $\theta=\pi/2$, ${\bm B}$ is purely along the ${\bm e}_\theta$ direction. Their strengths both increase toward short distances and become significantly smaller when the distance is about twice the shielding core size. This indicates that molecules experience the strongest effect of this mutual gauge field when their relative distance lies in a sphere shell just outside of the shielding core. We note that the average inter-molecular distance is indeed in this regime for the typical density of current experiments \cite{molecularBEC2024}. To quantify the strength of the synthetic magnetic field, we calculate the total magnetic flux through the equator outside of the shielding core, i.e. 
\begin{equation}
\Phi=2\pi\int_{r_\text{c}}^\infty rdr {\bm B}(r,\theta=\pi/2)\cdot{\bm e}_z.
\end{equation}
This result of $\Phi$ is shown in Fig. \ref{gauge}(b) as a function of $\Omega/\Delta$. It shows that the magnetic flux can easily reach the order of one flux quanta, indicating that it can produce a significant effect by altering the trajectory of relative motion between two molecules.  

Before ending this section, let us remark that the situation is very different if the microwave is pure $\pi$ polarized instead of $\sigma^+$ polarized. The $\pi$ polarized microwave only couples $\ket{g}$ to $\ket{e_0}$. We can repeat the analysis above by defining $\hat{U}(\phi)$ as
\begin{equation}
{\hat U}(\phi)=\exp[i \phi
(\hat N_{e_{-1}}-\hat N_{e_1}
)].
\end{equation}
This also leads to ${\bm A}({\bm r})=A_\phi(r,\theta){\bm e}_\phi$ and
\begin{equation}
A_\phi(r,\theta)=\frac{-\hbar}{r\sin\theta}\bra {\zeta(r,\theta)}\big(\hat N_{e_{-1}}-\hat N_{e_1}\big)\ket{\zeta(r,\theta)}. 
\end{equation}
However, with $\pi$ polarized microwave only, the system possesses the time-reversal symmetry, which ensures that $A_\phi(r,\theta)$ is always zero. Hence, $\sigma$ polarized microwave is essential for generating the synthetic mutual gauge field. 

If the microwave mixes $\sigma^+$ polarization with $\pi$ or $\sigma^{-}$ polarization, there is no such analytical expression of $\hat{U}$ that can eliminate all phase dependence in $\hat{W}({\bm r})$. However, one can always straightforwardly diagonalize $\hat{W}(\bf r)$ to obtain a complex eigenstate $\ket{\zeta({\bm r})}$ corresponding to the microwave-shielded potential. We can further compute ${\bm A}({\bm r})$ as $i\hbar\bra{\zeta({\bm r})}\nabla_{\bm r}\ket{\zeta({\bm r})}$. In this case, the resulting synthetic magnetic field no longer obeys cylindrical symmetry, leading to a distorted solenoid. However, the general physical picture should not change. We will leave the more detailed calculation for future investigation. 

\begin{figure}[t!]
    \centering
    \includegraphics[width=0.35\textwidth]{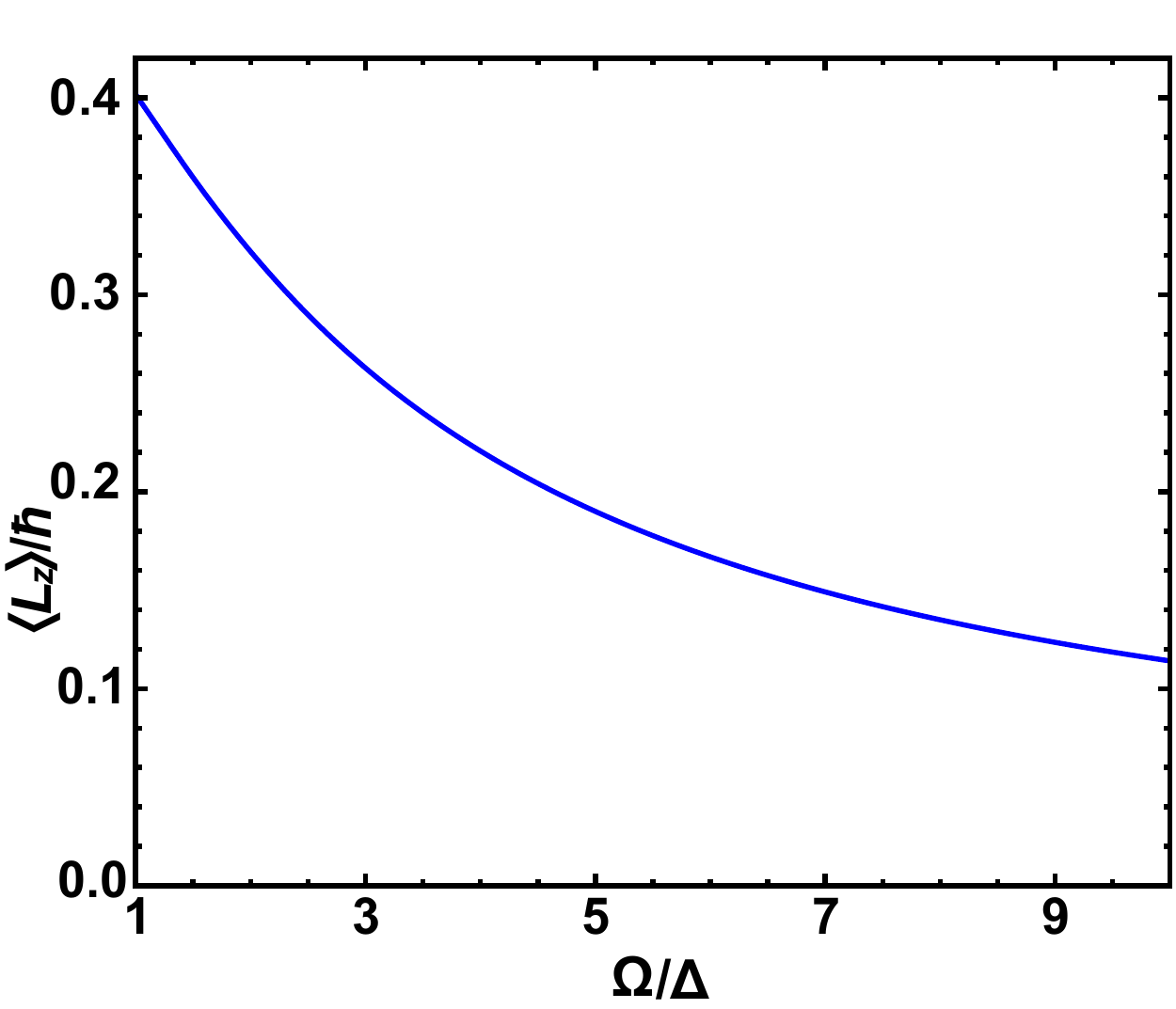}
    \caption{The relative angular momentum $\langle L_z\rangle$ for the ground state of two molecules in a two-dimensional harmonic trap as a function of $\Omega/\Delta$. As an example, we consider the system of two NaCs molecules. We set the harmonic trap frequency $\omega_{\rm ho}$ as $(2\pi)1{\rm MHz}$ and $\Delta$ as $\hbar(2\pi)10{\rm MHz}$. }
    \label{angular-momentum}
\end{figure}

\textit{Two-Body Problem.} Now, we discuss the physical consequence of this synthetic gauge field. The emergent magnetic field around each molecule is very similar to the magnetosphere around our Earth, which bends the trajectory of charged particles in the solar wind, except for those coming toward the two poles. By similar physical intuition, the dominant physical effect is to generate the mutual rotation between two molecules, especially when two molecules approach each other toward the shielding core in the $xy$ plane. To illustrate this effect, we solve a two-dimensional 
problem with two molecules restricted in the $xy$ plane and confined by a two-dimensional 
harmonic trap. We calculate the relative angular momentum for the ground state as
$\langle L_z\rangle=- i\hbar\int d{\bm r} \langle \Psi({\bm r})|\frac{ \partial }{\partial \phi} |\Psi({\bm r})\rangle$, where ${\bm r}$ now refers to the relative position in the $xy$ plane. As shown in Fig.~\ref{angular-momentum}, a non-zero relative angular momentum indeed appears.
The magnitude of this angular momentum is
 of the order of $\hbar$, consistent with the amount of the flux outside of the shielding core shown in Fig.~\ref{gauge}(b). 

The presence of $\sigma^+$ microwave breaks the time-reversal symmetry on the internal states. However, the finite relative angular momentum breaks the time-reversal symmetry in the spatial coordinate space. Hence, the two-body calculation reveals that the mutual gauge field is similar to spin-orbit coupling, which transmits the effect of time-reversal symmetry breaking from the internal space to the spatial coordinate space. In this sense, these two effects are similar. However, as we have emphasized at the beginning, conventional spin-orbit coupling is a single-particle physics, while this mutual gauge field occurs in the relative motion between any two molecules, which is an intrinsic interacting physics.  With this intuition, we expect to detect the manifestation of this gauge field in many-body systems by looking at the breaking of time-reversal symmetry in terms of the spatial motion of molecules. For instance, in a Bose-Einstein condensate of microwave-shielded molecules, the critical rotational frequencies for generating vortices should be different between clockwise and anti-clockwise rotation. One should also expect to observe the classical Hall effect in this system. Measuring critical rotation frequency and classical Hall effect has been realized in atomic condensate before and can be straightforwardly applied to molecular condensate~\cite{Madison2000,Beeler2013}.                                      

\begin{table*}[t]
        \centering
        \begin{tabular}{|c|c|c|}
        \hline
        &Similarity & Difference\\
        \hline
Liquid Helium       & Hardcore repulsion;
        & With versus without gauge field   \\ 
 & Interaction range comparable      & Isotropic versus anisotropic interaction \\
           & to interparticle spacing &    \\
        \hline
      & Gauge field emerges & Single particle gauge field \\
         Cold Atoms with  Synthetic Gauge Field & because of similar mechanism: & versus \\
           & internal states twist spatially & mutual gauge field  \\
        \hline
        Fractional Quantum Hall Effect & Repulsive interaction; 
        & Flux attachment versus  \\
        &  Presence of a mutual gauge field
        & solenoid attachment  \\
        \hline
        \end{tabular}
        \caption{The comparison of similarity and difference between microwave-shielded polar molecules and other three quantum systems studied extensively in the past.   }
        \label{comparision}
        \label{gauge-theory}
    \end{table*}

\textit{Concluding Remarks.} Having discussed the qualitative effect of the gauge field, here we emphasize that a quantitative study of such quantum many-body systems is highly challenging, to the extent that there is so far no protocol to write down reliably a Hamiltonian or Lagrangian. First of all, the mutual gauge field occurs between any pairs of molecules and couples to all relative coordinates, but we cannot write the kinetic energy term in the relative coordinates of all pairs since
\begin{equation}
\sum\limits_{i=1}^{N}\frac{\hbar^2}{2m}\nabla_{{\bm r}_i}^2\neq \sum\limits_{i\neq j}\frac{\hbar^2}{m}\nabla^2_{{\bm r}_{ij}}.
\end{equation}
This prevents including the mutual gauge field in the first quantization form of the many-body Hamiltonian. In the fractional quantum Hall case, the flux attachment is implemented in the field theory form by the Chern-Simons gauge theory. However, so far, it is also not clear how to attach a solenoid in a field theory approach. 

 To highlight the potential impact of this finding on future quantum simulation, we present in Table \ref{comparision} the similarities and differences between microwave-shielded polar molecules and the other three extensively studied quantum many-body systems. First, in comparison to liquid helium, both systems exhibit hard-core repulsion, with an interaction range comparable to the inter-particle distance \cite{Jin2024}. The long-range nature of the interaction can lead to a supersolid phase for bosons or a Wigner crystal phase for fermions. Investigating the unconventional properties induced by the emergent gauge field in polar molecules within the supersolid or Wigner crystal phase is worthwhile. Secondly, when compared to cold atoms with synthetic gauge fields, both systems show effects of spin-orbit coupling as discussed earlier. The key difference is that in cold atoms, the gauge field pertains to single-particle physics, whereas it is part of the interaction in polar molecules. Spin-orbit coupling is known to be an essential ingredient for topological physics, making it intriguing to explore whether interaction-driven topology will emerge in this system. Thirdly, compared to the fractional quantum Hall effect, both systems feature repulsive interactions and mutual gauge fields. The angular momentum displayed in Fig. \ref{angular-momentum} is also sufficiently large to ensure that the system transitions into the fractional quantum Hall regime. However, the difference lies in the structure of the mutual gauge field. In the fractional quantum Hall case, a quantized flux is locally attached to each electron \cite{Nagaosa}, whereas, in the polar molecule case, the magnetic field has an extended spatial distribution around each molecule. This enhances the spatial fluctuation of the gauge field as molecules move, potentially resulting in novel types of strongly correlated phases in the fractional quantum Hall context.

{\bf Supplementary:  Derivation of Eq.~(\ref{vdd}).}  Eq.~(\ref{vdd}) is the expression of the dipolar interaction in the rotating frame under the rotating-wave approximation (RWA). This expression was first derived by Ref.~\cite{Deng2023}, and we re-derived this expression using a slightly different approach and with our notation, to be self-contained.  

	As discussed in the main text, we first project the expression in Eq.~(\ref{vsdd}) onto the subspace spanned by the two-molecule states \( |\alpha_1;\alpha_2\rangle  \) (\( \alpha_{1,2} = g, e_{0,\pm 1} \)) and obtain:
\begin{eqnarray}
\hat{H}_\text{int}({\bm r}) = \sum_{\alpha_1,\alpha_2,\alpha_1^\prime,\alpha_2^\prime} h_{\alpha_1,\alpha_2}^{\alpha_1^\prime,\alpha_2^\prime} ({\bm r}) |\alpha_1;\alpha_2\rangle \langle \alpha_1^\prime;\alpha_2^\prime|.\label{a1}
\end{eqnarray}
The matrix element \( h_{\alpha_1,\alpha_2}^{\alpha_1^\prime,\alpha_2^\prime} (\bm r) \) is given by:
\begin{eqnarray}
&&h_{\alpha_1,\alpha_2}^{\alpha_1^\prime,\alpha_2^\prime} (\bm r)\nonumber\\
& =& \langle \alpha_1;\alpha_2 | \frac{1}{4\pi \epsilon_0 r^3} \left[ \hat{\bm{d}}_1 \cdot \hat{\bm{d}}_2 - \frac{3}{r^2} (\hat{\bm{d}}_1 \cdot \bm{r}) (\hat{\bm{d}}_2 \cdot \bm{r}) \right] | \alpha_1^\prime; \alpha_2^\prime \rangle.\nonumber\\
\label{a2}
\end{eqnarray}
By applying the RWA, we ignore the off-resonant terms on the right-hand side of Eq.~(\ref{a1}) and only keep the terms satisfying \( E_{\alpha_1} + E_{\alpha_2} = E_{\alpha_1^\prime} + E_{\alpha_2^\prime} \), where \( E_g = 0 \) and \( E_{e_{0,\pm 1}} = 2\Lambda \) are the bare energies of the one-molecule internal states \( |g\rangle \) and \( |e_{0,\pm 1}\rangle \), respectively.
Under the RWA, the dipolar interaction is given by:
\begin{eqnarray}
\hat{H}_\text{int}({\bm r}) &=& \sum_{\alpha_1, \alpha_2, \alpha_1^\prime, \alpha_2^\prime} h_{\alpha_1, \alpha_2}^{\alpha_1^\prime, \alpha_2^\prime} (\bm r) \delta_{E_{\alpha_1}+E_{\alpha_2}, E_{\alpha_1^\prime}+E_{\alpha_2^\prime}}\nonumber\\
&&\hspace{1.5cm}
\times |\alpha_1;\alpha_2 \rangle \langle \alpha_1^\prime; \alpha_2^\prime|.\label{a3}
\end{eqnarray}
Furthermore, we note Eq. \ref{a3} is invariant under the rotating wave transformation.  

Finally, we notice that dipole operator ${\hat {\bm d}}_j$ ($j=1,2$) of molecule $j$ can be expressed as ${\hat {\bm d}}_j=d\sqrt{4\pi/3}[{\hat d}_j^{(0)}{\bf e}_z+{\hat d}_j^{(+)}{\bf e}_++{\hat d}_j^{(-)}{\bf e}_-]$, where $d$ is the dipole moment and ${\bf e}_\pm=(\mp{\bf e}_x+i{\bf e}_y)/\sqrt{2}$, with ${\bf e}_{x,y,z}$ being the unit vectors along the $x,y,z$ axis, respectively. Furthermore, in the subspace spanned by $|g\rangle_j$ and $|e_{0,\pm 1}\rangle_j$ ($j=1,2$) 
the operators ${\hat d}_j^{(0,+)}$ only have the following non-zero matrix elements:
\begin{eqnarray}
_j\!\bra{g}{\hat d}_j^{(0)}\ket{e_0}_j&=&_j\!\bra{e_0}{\hat d}_j^{(0)}\ket{g}\!_j=\frac{1}{\sqrt{4\pi}};\\
_j\!\bra{e_1}{\hat d}_j^{(+)}\ket{g}\!_j&=&\frac{1}{\sqrt{4\pi}};\\
_j\!\bra{g}{\hat d}_j^{(+)}\ket{e_{-1}}_j&=&-\frac{1}{\sqrt{4\pi}},
\end{eqnarray}
and the non-zero matrix elements of ${\hat d}_j^{(-)}$ in this subspace can be derived via the relation ${\hat d}_j^{(-)}=-{\hat d}_j^{(+)\dagger}$. Using these coefficients we can derive the explicit expressions for $h_{\alpha_1,\alpha_2}^{\alpha_1^\prime,\alpha_2^\prime} (\bm r)$ of Eq.~(\ref{a2}). We finally arrive at Eq.~(\ref{vdd}) by substituting these expressions into the right-hand side of Eq.~(\ref{a3}).

{\it Acknowledgement.} We would like to thank Tao Shi, Ren Zhang, Yadong Wu, and Zeqing Wang for their discussions and help. This work is supported by the National
Key R\&D Program of China 2022YFA1405301 (PZ and RQ), 2023YFA1406702 (HZ),   
the Innovation Program for
Quantum Science and Technology 2021ZD0302005 (HZ), 2023ZD0300700 (PZ),
the XPLORER Prize (HZ), 
the National Natural Science Foundation of China (12022405 (RQ), 11774426 (RQ), and U23A6004 (HZ)), and
the Beijing Natural Science Foundation (Z180013 (RQ)).

{\bf Code availability.} Codes for numerical calculations and figures are available at \url{https://github.com/xubei0903-phy/ShieldedMolecule-gauge}.

\bibliographystyle{apsrev4-2}

\bibliography{ref}

\end{document}